\documentclass[]{spie}  
\usepackage[]{graphicx}
\usepackage{amsmath}
\usepackage{multirow}
\usepackage[pdfpagemode=None,pdfstartview=FitH,pdfview=FitH,colorlinks=true,pdftitle=Predicting\ Chroma\ from\ Luma\ with\ Frequency\ Domain\ Intra\ Prediction,pdfauthor=Nathan\ E.\ Egge]{hyperref}

\def\sgn{\mathop{\rm sgn}}

\title{Predicting Chroma from Luma with Frequency Domain Intra Prediction}

\author{Nathan E. Egge and Jean-Marc Valin
\skiplinehalf
Mozilla, Mountain View, USA
\skiplinehalf
Xiph.Org Foundation
}

\authorinfo{Copyright 2014-2015 Mozilla Foundation. This work is licensed under \href{https://creativecommons.org/licenses/by/3.0/}{CC-BY 3.0}.\\Send correspondence to Nathan E. Egge $<$\href{mailto:negge@xiph.org}{negge@xiph.org}$>$.}

\begin{document} 
  \maketitle 

\begin{abstract}
This paper describes a technique for performing intra prediction of the chroma
 planes based on the reconstructed luma plane in the frequency domain.
This prediction exploits the fact that while RGB to YUV color conversion has
 the property that it decorrelates the color planes globally across an image,
 there is still some correlation locally at the block level\cite{LeeCho09}.
Previous proposals compute a linear model of the spatial relationship between
 the luma plane (Y) and the two chroma planes (U and V)\cite{JCTVCB021}.
In codecs that use lapped transforms this is not possible since transform
 support extends across the block boundaries\cite{Tran2003} and thus
 neighboring blocks are unavailable during intra-prediction.
We design a frequency domain intra predictor for chroma that exploits the same
 local correlation with lower complexity than the spatial predictor and which
 works with lapped transforms.
We then describe a low-complexity algorithm that directly uses luma coefficients
 as a chroma predictor based on gain-shape quantization and band
 partitioning.
An experiment is performed that compares these two techniques inside the
 experimental Daala video codec and shows the lower complexity algorithm to be
 a better chroma predictor.
\end{abstract}



\section{INTRODUCTION}
\label{sec:intro}  

\begin{figure}
\begin{center}
\begin{tabular}{c}
\includegraphics[natwidth=934,natheight=594,width=4in]{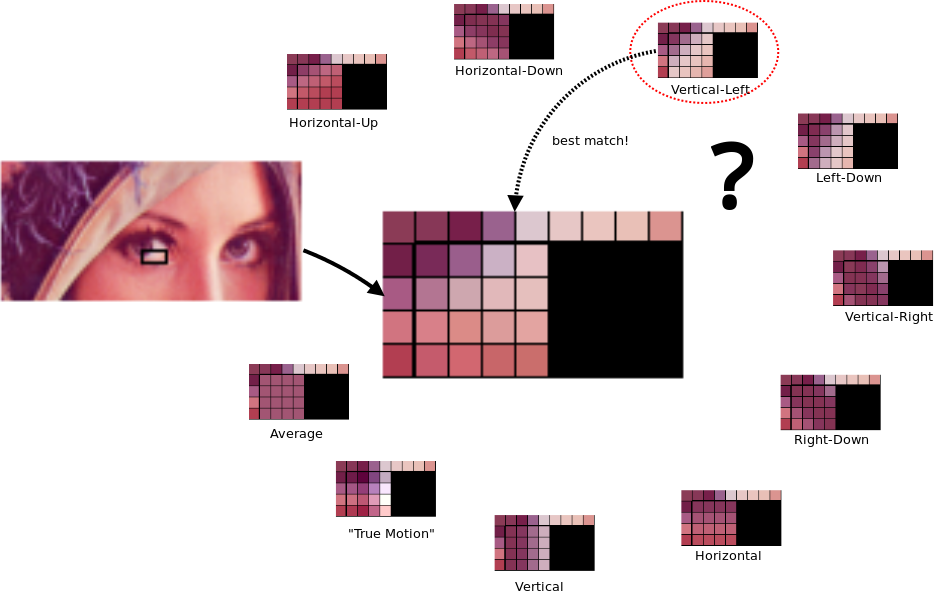}
\end{tabular}
\end{center}
\caption[example]{\label{fig:webp} The 10 intra-prediction modes for 4x4 blocks in
 WebM (VP8).}
\vspace{15pt}
\end{figure}

Still image and video codecs typically consider the problem of intra-prediction
 in the spatial domain.
A predicted image is generated on a block-by-block basis using the previously
 reconstructed neighboring blocks for reference, and the residual is encoded
 using standard entropy coding techniques.
Modern codecs use the boundary pixels of the neighboring blocks along with a
 directional mode to predict the pixel values across the target block (e.g.,
 AVC, HEVC, VP8, VP9, etc.).
These directional predictors are cheap to compute (often directly copying pixel
 values or applying a simple linear kernel), exploit local coherency (with low
 error near the neighbors) and predict hard to code features (extending sharp
 directional edges across the block).
In Figure \ref{fig:webp} the ten intra-prediction modes of WebM are shown for a
 given input block based on the 1 pixel boundary around that block.

\begin{figure}[h!]
\begin{center}
\begin{tabular}{c}
\includegraphics[natwidth=1376,natheight=646,width=3.5in]{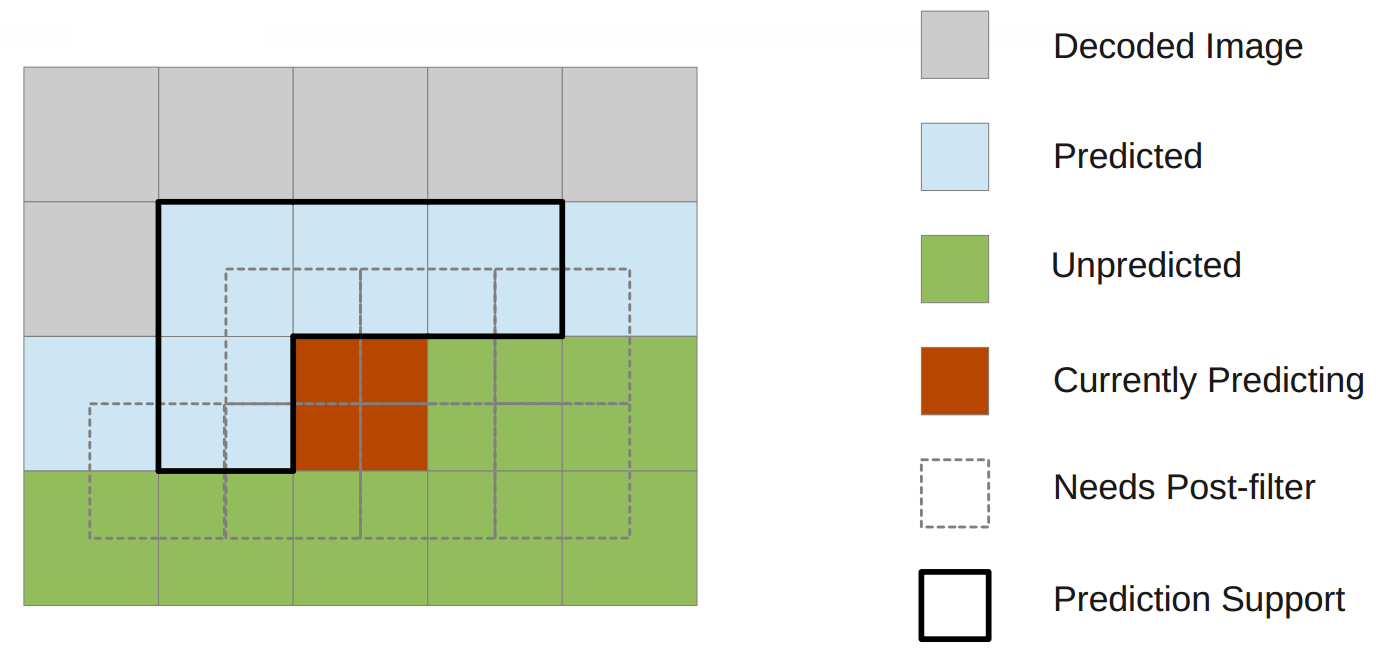}
\end{tabular}
\end{center}
\caption[example]{\label{fig:decode} State of blocks in the decode pipeline of
 a codec using lapped transforms. Immediate neighbors of the target block
 (bold lines) cannot be used for spatial prediction as they still require
 post-filtering (dotted lines).}
\end{figure}

In codecs that use lapped transforms these techniques are not applicable (e.g.,
 VC-1, JPEG-XR, Daala, etc.).
The challenge here is that the neighboring spatial image data is not available
 until {\em after} the target block has been decoded and the appropriate
 unlapping filter has been applied across the block boundaries.
Figure \ref{fig:decode} shows the decode pipeline of a codec using lapped
 transforms with a single block size.
The support used in spatial intra prediction is exactly the region that has not
 had the unlapping post-filter applied.
Note that the pre-filter has the effect of decorrelating the image along block
 boundaries so that the neighboring pixel values before unlapping are
 particularly unsuitable for use in prediction.


Work has been done to use intra prediction with lapped transforms.
Modifying AVC, de Oliveira and Pesquet showed that it was possible to use the
 boundary pixels just outside the lapped region to use 4 of the 8 directional
 intra prediction modes with lapped transforms\cite{oliv2011}.
The work of Xu, Wu and Zhang considers prediction as a transform and proposes
 a frequency domain intra prediction method using non-overlapped
 blocks\cite{xuwu2009}.
An early experiment with the Daala video codec extended this idea using
 machine learning to train sparse intra predictors\cite{DaalaDemo2}.
However this technique is computationally expensive (4 multiplies per
 coefficient) and not easily vectorized.

A promising technique was proposed by Lee and Cho to predict the chroma
 channels using the spatially coincident reconstructed luma
 channel\cite{LeeCho09}.
This was formally proposed for use in HEVC by Chen et al\cite{JCTVCB021}
 however was ultimately not selected due to the increased encoder and decoder
 running time of 20-30\%.
We propose a similar technique that adapts the spatial chroma-from-luma
 intra prediction for use with frequency-domain coefficients.
We call this algorithm frequency-domain chroma-from-luma (FD-CfL).

More recently, work on the Daala video codec has included replacing scalar
 quantization with gain-shape quantization\cite{valin2015spie}.
We show that when prediction is used with gain-shape quantization, it is
 possible to design a frequency-domain chroma-from-luma predictor without
 the added encoder and decoder overhead.
An experimental evaluation between FD-CfL and the proposed PVQ-CfL algorithm
 shows this reduction in complexity comes with no penalty to quality and
 actually provides an improvement at higher rates.

\section{CHROMA FROM LUMA PREDICTION}
\label{sec:chroma}

In spatial-domain chroma-from-luma, the key observation is that the local
 correlation between luminance and chrominance can be exploited using a linear
 prediction model.
For the target block, the chroma values can be estimated from the reconstructed
 luma values as
\begin{align}
C(u,v) & = \alpha\cdot L(u,v) + \beta
\end{align}
where the model parameters $\alpha$ and $\beta$ are computed as a linear
 least-squares regression using $N$ pairs of spatially coincident luma and
 chroma pixel values along the boundary:
\begin{align}
\alpha & = \frac{N\cdot\displaystyle\sum_i L_i\cdot C_i - \displaystyle\sum_i L_i\displaystyle\sum_i C_i}{N\cdot\displaystyle\sum_i L_i\cdot L_i - \left(\displaystyle\sum_i C_i\right)^2}, & \beta & = \frac{\displaystyle\sum_i C_i -\alpha\cdot\displaystyle\sum_i L_i}{N}.
\label{eqn:fit}
\end{align}

When $\alpha$ and $\beta$ are sent explicitly in the bitstream, the pairs
 $(L_i,C_i)$ are taken from the original, unmodified image.
However, the decoder can also compute the same linear regression using its
 previously decoded neighbors and thus $\alpha$ and $\beta$ can be
 derived {\em implicitly} from the bitstream.
Additional computation is necessary when the chroma plane is subsampled (e.g.,
 4:2:0 and 4:2:2 image data) as the luma pixel values are no longer coincident
 and must be resampled.
In the next section we adapt the algorithm to the frequency-domain and show
 that this issue does not exist at most block sizes.

\subsection{EXTENSION TO FREQUENCY-DOMAIN}
\label{sec:alg}

In codecs that use lapped transforms, the reconstructed pixel data is not
 available.
However the transform coefficients in the lapped frequency domain are the
 product of two linear transforms: the linear pre-filter followed by the linear
 forward DCT.
Thus the same assumption of a linear correlation between luma and chroma
 coefficients holds.
In addition, we can take advantage of the fact that we are in the frequency
 domain to use only a small subset of coefficients when computing our model.

The chroma values can then be estimated using frequency-domain chroma-from-luma
 (FD-CfL):
\begin{align}
C_{DC} &= \alpha_{DC}\cdot L_{DC} + \beta_{DC} \\
C_{AC}(u,v) &= \alpha_{AC}\cdot L_{AC}(u,v)\label{eqn:cfl_ac}
\end{align}
where the $\alpha_{DC}$ and $\beta_{DC}$ are computed using the linear
 regression in Equation \ref{eqn:fit} with the DC coefficients of the three
 neighboring blocks: up, left and up-left.
When estimating $C_{AC}(u,v)$ we can omit the constant offset $\beta_{AC}$
 as we expect the AC coefficients to be zero mean.
Additionally, we do not include all of the AC coefficients from the same three
 neighboring blocks when computing $\alpha_{AC}$.

\begin{figure}
\begin{center}
\begin{tabular}{c c c}
\includegraphics[natwidth=200,natheight=200,width=1.75in]{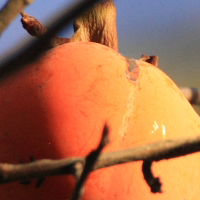}
&
\includegraphics[natwidth=200,natheight=200,width=1.75in]{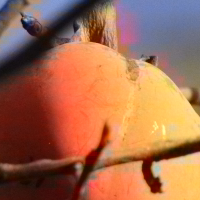}
&
\includegraphics[natwidth=200,natheight=200,width=1.75in]{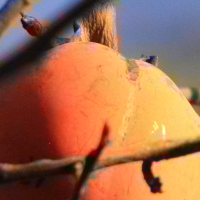}
\\
(a) & (b) & (c)
\end{tabular}
\end{center}
\caption[example]{\label{fig:comp} Comparison of (a) the original uncompressed
 image with composite images of (b) reconstructed luma and predicted chroma
 using experimental Daala intra modes and (c) reconstructed luma and predicted
 chroma using FD-CfL.}
\end{figure}

It is sufficient to use the three lowest AC coefficients from the neighboring
 blocks.
This means that the number of input pairs $N$ is constant regardless of the
 size of chroma block being predicted.
Moreover, the input AC coefficients have semantic meaning: we use the
 strongest horizontal, vertical and diagonal components.
This has the effect of preserving features across the block as edges are
 correlated between luma and chroma, see the fruit and branch edges in
 Figure \ref{fig:comp} (c).

\subsection{TIME-FREQUENCY RESOLUTION SWITCHING}
\label{sec:tf}
When image data is 4:4:4 or 4:2:0, the chroma and luma blocks are aligned so
 that the lowest 3 AC coefficients describe the same frequency range.
In codecs that support multiple block sizes (or that
 support 4:2:2 image data) it is the case that the luma blocks and the chroma
 blocks are not aligned.
For example, in the Daala video codec the smallest block size supported is 4x4.
In 4:2:0, when an 8x8 block of luma image data is split into four 4x4 blocks,
 the corresponding 4x4 chroma image data is still coded as a single 4x4 block.

\begin{figure}[h]
\begin{center}
\begin{tabular}{c c c}
\includegraphics[natwidth=327,natheight=138,width=2in]{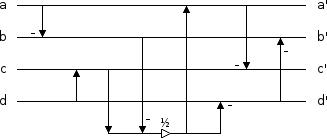}
&

&
\includegraphics[natwidth=799,natheight=237,width=4in]{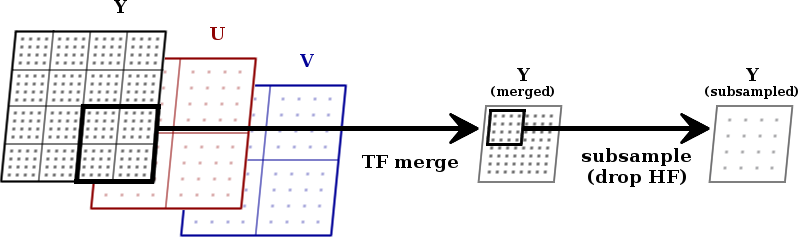}\\
(a) & & (b)
\end{tabular}
\end{center}
\caption[example]{\label{fig:tf} (a) Circuit diagram of the 2x2 TF-merge
 algorithm that uses 7 adds and 1 shift.
(b) Using TF-merge to convert four 4x4 frequency
 domain luma blocks (Y) into a single 8x8 frequency domain block for use with
 TD-CfL when image data is 4:2:0.  Note that only the lower frequency (LF)
 portion is actually necessary and thus only that portion of the TF-merge need
 be computed.}
\end{figure}

This is a problem for FD-CfL as it requires the reconstructed luma
 frequency-domain coefficients to cover the same spatial extent.
In Daala this is overcome by borrowing a technique from the Opus audio
 codec\cite{valin2013high}.
Using Time-Frequency resolution switching (TF) it is possible to trade off
 resolution in the spatial domain for resolution in the frequency domain.
Here the four 4x4 luma blocks are {\em merged} into a single 8x8 block with
 half the spatial resolution and twice the frequency resolution.
We apply a 2x2 Hadamard transform to corresponding transform coefficients in
 four 4x4 blocks to merge them into a single 8x8 block.
The low frequency (LF) coefficients are then used with FD-CfL, see
 Figure \ref{fig:tf}.

\subsection{COMPLEXITY COMPARISON}

Both the spatial and frequency domain chroma-from-luma techniques have the
 property that once the model parameters $\alpha$ and $\beta$ have been
 determined the entire chroma block can be computed with one multiply and one
 add per pixel.
This is far better than the frequency domain intra prediction used to predict
 the luma plane in Daala which uses 4 multiplies and 4 adds\cite{DaalaDemo2}.
An important question to answer is how does the computational complexity of
 the model fitting step in FD-CfL compare to the HEVC proposal\cite{JCTVCB021}.
Note that the HEVC proposal required resampling the entire luma block for 4:2:0
 image data, a process which is roughly as expensive as a TF-merge and
 is necessary for all luma block sizes (not just at the smallest block size).

\begin{table}[h]
\begin{center}
\begin{tabular}{|c|c|c|c|c|}
\hline
\multirow{2}{*}{Block Size} & \multicolumn{2}{|c|}{Spatial Domain} & \multicolumn{2}{|c|}{Freq. Domain} \\
\cline{2-5}
 & Mults & Adds & Mults & Adds \\
\hline
$N\times N$ & $4*N+2$ & $8*N+3$ & $2*12+5$ & $4*12+5$ \\
\hline
$4\times 4$ & 18 & 35 & 29 & 53 \\
\hline
$8\times 8$ & 34 & 67 & 29 & 53 \\
\hline
$16\times 16$ & 66 & 131 & 29 & 53 \\
\hline
\end{tabular}
\end{center}
\caption[example]{\label{tbl:comp} Comparison of cost to fit the model in
 Equation \ref{eqn:fit} using spatial and frequency domain chroma-from-luma.}
\end{table}

To answer this question, Table \ref{tbl:comp} is provided which compares just
 the cost of fitting the chroma-from-luma model.
An interesting feature of the FD-CfL algorithm is that the cost to fit the
 model is independent of the block size, as we only consider a small portion
 of the frequency domain coefficients.
In the frequency domain we use 12 $(L_i,C_i)$ pairs (the 4 LF coefficients
 from the 3 neighboring blocks), versus $2*N$ pairs in the spatial domain ($N$
 from each of the bordering left and up blocks).
This means that for all chroma block sizes larger than 4x4, model fitting in
 the frequency domain is actually cheaper than in the spatial domain.

\section{GAIN-SHAPE QUANTIZATION}
\label{sec:pvq}

Most modern video and still image codecs use scalar quantization as a ``lossy''
 way of reducing the amount of information needed to code a block.
After the block coefficients have been transformed into the frequency-domain,
 they are each quantized to an integer index which is entropy coded.
In the decoder the transform coefficients are reconstructed by reversing the
 quantization.
For a coefficient $C_i$ and quantization step size $Q_i$, the quantization
 index $\gamma_i$ and reconstructed coefficient $\hat{C_i}$ can be found by
\begin{align}
\gamma_i & = \lfloor C_i / Q_i\rfloor \\
\hat{C_i} & = \gamma_i\cdot Q_i
\end{align}
Note that when the quantization step size is large (e.g., at low rates) the
 smaller, high frequency coefficients go to zero.
While good for compression (many codecs will run-length encode a string of
 zeros), this has the effect of low-passing the block and removing much of the
 texture from the image.

An alternative to scalar quantization is to use vector quantization (VQ).
Here the quantization index $\gamma$ no longer represents a single coefficient,
 but rather an entire vector of coefficients.
The idea is to take, for example, the entire block of coefficients and treat
 them as an $n$-dimensional vector.
Quantization then amounts to finding the index $\gamma$ of the nearest codeword
 ($n$-dimensional vector) in a possibly infinite VQ-codebook.
The density of codewords in the codebook around the $n$-dimensional vector we
 are quantizing dictates the quantization error.
However, it has been shown that even for a fixed set of input vectors,
 designing an optimal VQ-codebook is an NP-hard problem.
Moreover, searching for the optimal quantization index $\gamma$ requires
 computing the distance between the input vector and every VQ-codeword to
 select the closest.

In the Daala video codec we use gain-shape quantization\cite{DaalaDemo6}.
A vector of coefficients ${\bf x}$ is separated into two intuitive components:
 its magnitude ({\em gain}) and its direction ({\em shape}).
The gain $g=\left\|{\bf x}\right\|$ represents how much energy is contained in the
 block, and the shape ${\bf u}={\bf x}/\left\|{\bf x}\right\|$ indicates where that energy is
 distributed among the coefficients.
The gain is then quantized using scalar quantization, while the shape is
 quantized by finding the nearest VQ-codeword in an algebraically defined
 codebook.
This has the advantage of not needing to explicitly store the VQ-codebook in
 the decoder as well as allowing the encoder to search only a small set of
 VQ-codewords around the input vector.
Given the gain quantization index $\gamma_g$, the shape vector quantization
 index $\gamma_u$ and an implicitly defined VQ-codebook $CB$, the reconstructed
 gain $\hat{g}$ and shape $\hat{{\bf u}}$ can be found by
\begin{align}
\hat{g} & = \gamma_g \cdot Q \\
\hat{{\bf u}} & = CB[{\gamma_u}]
\end{align}
and reconstructed coefficients $\hat{{\bf x}}$ are thus
\begin{align}
\hat{{\bf x}} & = \hat{g}\cdot \hat{{\bf u}}
\end{align}

By explicitly signaling the amount of energy in a block, and roughly where that
 energy is located, gain-shape quantization is texture preserving.
Because the algebraic codebook used in Daala is based on the pyramid vector
 quantizer described by Fisher\cite{Fisher1986}, this technique is referred to
 as Perceptual Vector Quantization (PVQ).
A complete description of PVQ usage in Daala and its other advantages over
 scalar quantization is outside the scope of this paper and and is described
 in detail by Valin\cite{valin2015spie}.

\subsection{PREDICTION WITH PVQ}

In block based codecs, both intra- and inter-prediction can often construct a
 very good predictor for the block that will be decoded next.
In the encoder, this predicted block is typically subtracted from the input
 image and the residual is transformed to the frequency domain, quantized and
 entropy coded.
When the transform is linear, as is the case with codecs based on lapped
 transforms, this is equivalent to transforming the predictor and computing
 the difference in the frequency domain.
However, if one were to simply quantize the frequency domain residual using PVQ,
 the texture preservation property described in the previous section would be
 lost.
This is because the energy that would be preserved is no longer that of the
 block being coded, but instead the gain represents how much the image is
 different from its predictor.
In Daala, this is avoided by explicitly {\em not} computing a residual, but
 instead extracting another intuitive parameter in gain-shape quantization.

\begin{figure}
\begin{center}
\begin{tabular}{c c}
\includegraphics[natwidth=650,natheight=530,width=3in]{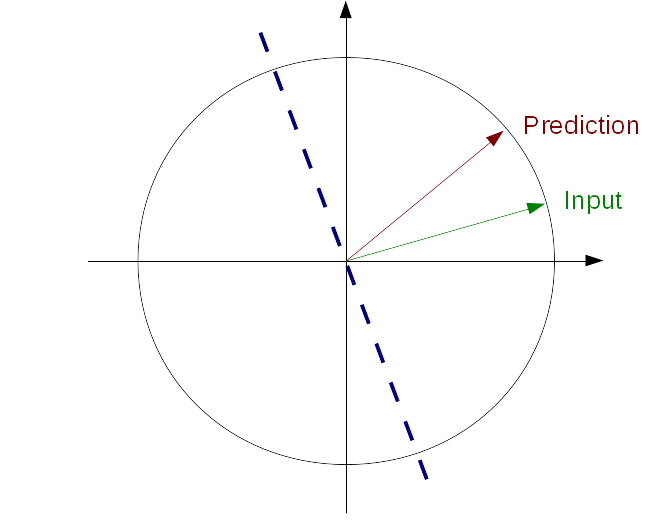} &
\includegraphics[natwidth=650,natheight=530,width=3in]{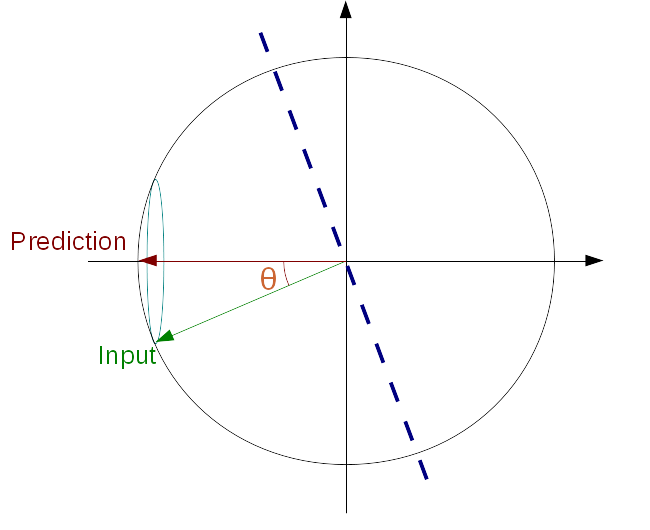} \\
(a) & (b)
\end{tabular}
\end{center}
\caption[pvq]{\label{fig:pvq} (a) A Householder reflection plane is computed
 that aligns the prediction vector so that its largest component is along an
 axis.  (b) The input vector is reflected across the plane and the angle
 $\theta$ is computed and coded using scalar quantization. The axis on which
 the predictor lies is eliminated, the remaining $n-1$ dimensions are coded
 using PVQ.}
\end{figure}

Ideally, when the predictor is good we would like the cost of coding the gain
 and shape to be low.
That is, we would like the entropy of the symbols we code to be as small as
 possible.
We can achieve this and retain the energy preserving properties of PVQ by
 using a Householder reflection.
Considering the predictor as another $n$-dimensional vector, a reflection plane
 is computed that aligns the predictor with one of the axes in our
 $n$-dimensional vector space making all but one of the components in the
 predictor equal zero.
The encoder can then reflect the input vector ${\bf x}$ across this reflection
 plane in a way that is perfectly reproducible in the decoder, see
 Figure \ref{fig:pvq}.

Let ${\bf r}$ be the $n$-dimensional vector of predictor coefficients.  Then the
 normal to the reflection plane can be computed as
\begin{align}
{\bf v} & = \frac{{\bf r}}{\left\|{\bf r}\right\|}+s\cdot {\bf e}_m
\end{align}
where $s\cdot {\bf e}_m$ is the signed unit vector in the direction of the axis we
 would like to reflect ${\bf r}$ onto.
The input vector ${\bf x}$ can then be reflected across this plane by computing
\begin {align}
{\bf z} & = {\bf x}-2\frac{{\bf v}^T {\bf x}}{{\bf v}^T {\bf v}}{\bf v}
\end{align}
We can measure how well the predictor ${\bf r}$ matches our input vector
 ${\bf x}$ by computing the cosine of the angle $\theta$ between them as
\begin{align}
\cos\theta & = \frac{{\bf x}^T {\bf r}}{\left\|{\bf x}\right\| \left\|{\bf r}\right\|}
 = \frac{{\bf z}^T {\bf r}}{\left\|{\bf z}\right\| \left\|{\bf r}\right\|}
 = -s\frac{z_m}{\left\|{\bf z}\right\|}
\end{align}

We are free to choose any axis in our $n$-dimensional space and we select
 ${\bf e}_m$
 to be the dimension of the largest component of our prediction vector
 ${\bf r}$ and $s = \sgn(r_m)$.
Thus the largest component lies on the $m$-axis after reflection.
When the predictor is good, we expect that the largest component of ${\bf z}$
 will also be in the ${\bf e}_m$ direction and $\theta$ will be small.
If we code $\hat{\theta}$ using scalar quantization, we can remove the largest
 dimension of ${\bf z}$ and reduce the coding of ${\bf x}$ to a gain-shape
 quantization of the remaining $n-1$ coefficients where the gain has been
 reduced to $\sin\theta\cdot g$.
Given a predictor ${\bf r}$, the reconstructed coefficients $\hat{{\bf x}}$
 are computed as
\begin{align}
\hat{{\bf x}} = \hat{g}\big(-s\cdot\cos\hat{\theta}\cdot {\bf e}_m +
 \sin\hat{\theta}\cdot\hat{{\bf u}}\big)
\end{align}
When the predictor is poor, $\theta$ will be large and the reflection is
 unlikely to make things easier to code.
Thus when $\theta$ is greater than $90^{\circ}$ we code a flag and use PVQ with
 no predictor.
Conversely when ${\bf r}$ is exact, $\hat{\theta}$ is zero and no additional
 information needs to be coded.
In addition, because we expect ${\bf r}$ to have roughly the same amount of
 energy as ${\bf x}$, we can get additional compression performance by using
 $\left\|{\bf r}\right\|$ as a predictor for $g$:
\begin{align}
\hat{g} & = \gamma_g\cdot Q + \left\|{\bf r}\right\|
\end{align}

\subsection{CHROMA FROM LUMA USING PVQ PREDICTION}

Let us now return to the frequency-domain chroma-from-luma (FD-CfL) algorithm
 from Section \ref{sec:alg} and consider what happens when it is used with
 gain-shape quantization.
As an example, consider a 4x4 chroma block where the 15 AC coefficients are
 coded using gain-shape quantization with the FD-CfL predictor from
 Equation \ref{eqn:cfl_ac}.
The 15-dimensional predictor ${\bf r}$ is simply a linearly scaled vector of
 the coincident reconstructed luma coefficients:
\begin{align}
C_{AC}(u,v) = \alpha_{AC}\cdot L_{AC}(u,v)\implies {\bf r} = \alpha_{AC}\cdot\hat{{\bf x}}_L
\end{align}
Thus the shape of the chroma predictor ${\bf r}$ is exactly that of the
 reconstructed luma coefficients $\hat{{\bf x}}_L$ with one exception:
\begin{align}
\frac{{\bf r}}{\left\|{\bf r}\right\|} & =
 \frac{\alpha_{AC}\cdot\hat{{\bf x}}_L}{\left\|\alpha_{AC}\cdot\hat{{\bf x}}_L\right\|} =
 \sgn(\alpha_{AC})\frac{\hat{{\bf x}}_L}{\left\|\hat{{\bf x}}_L\right\|}
\end{align}
Because the chroma coefficients are sometimes inversely correlated with the
 coincident luma coefficients, the linear term $\alpha_{AC}$ can be negative.
In these instances the {\em shape} of $\hat{{\bf x}}_L$ points in exactly the
 wrong direction and must be flipped.

Moreover, consider what happens to the gain of ${\bf x}_C$ when it is predicted
 from ${\bf r}$.
The PVQ prediction technique assumes that
$\left\|{\bf r}\right\| = \alpha_{AC}\cdot\left\|\hat{{\bf x}}_L\right\|$ is a
 good predictor of $g_C = \left\|{\bf x}_C\right\|$.
Because $\alpha_{AC}$ for a block is learned from its previously decoded
 neighbors, often it is based on highly quantized or even zeroed coefficients.
When this happens, $\alpha_{AC}\cdot\left\|\hat{{\bf x}}_L\right\|$ is no
 longer a good predictor of $g_C$ and the cost to code
 $\left\|{\bf x}_C\right\|-\alpha_{AC}\cdot\left\|\hat{{\bf x}}_L\right\|$
 using scalar quantization is actually greater than the cost of just coding
 $g_C$ alone.

Thus we present a modified version of PVQ prediction that is used just for
 chroma-from-luma intra prediction called PVQ-CfL.
For each set of chroma coefficients coded by PVQ, the prediction vector
 ${\bf r}$ is exactly the coincident luma coefficients.
Note that for 4:2:0 video we still need to apply the Time-Frequency resolution
 switching (TF) described in Section \ref{sec:tf} to merge the reconstructed
 coefficients of 4x4 luma blocks to get the coincident predictor
 $\hat{{\bf x}}_L$ for the corresponding 4x4 chroma block.
We determine if we need to flip the predictor by computing the sign of the
 cosine of the angle between $\hat{{\bf x}}_L$ and ${\bf x}_C$:
\begin{align}
f & = \sgn(\hat{{\bf x}}_L^T {\bf x}_C)
\end{align}
A negative sign means the angle between the two is greater than $90^{\circ}$
 and flipping $\hat{{\bf x}}_L$ is guaranteed to make the angle less than
 $90^{\circ}$.

We then code $f$ using a single bit\footnotemark[3], and the gain
 $\hat{g}_C$ using scalar quantization with no predictor.
The shape quantization algorithm for ${\bf x}_C$ is unchanged except that
 ${\bf r} = f\cdot \hat{{\bf x}}_L$.
This algorithm has the advantage over FD-CfL of being both lower complexity
 (neither the encoder nor decoder need to compute a linear regression per block)
 and providing better compression (the chroma gain $g_C$ is never incorrectly
 predicted).
\footnotetext[3]{It is not strictly necessary to code a bit for $f$.  Instead
 the parameter $\alpha_{AC}$ could be found using least-squares regression and
 the sign extracted.  However, using a single bit to code $f$ is (1) better
 overall than relying on least-squares regression which can be wrong and (2)
 reduces the complexity significantly.}

\subsection{PVQ WITH FREQUENCY BANDS}

\begin{figure}
\begin{center}
\begin{tabular}{c}
\includegraphics[width=2in]{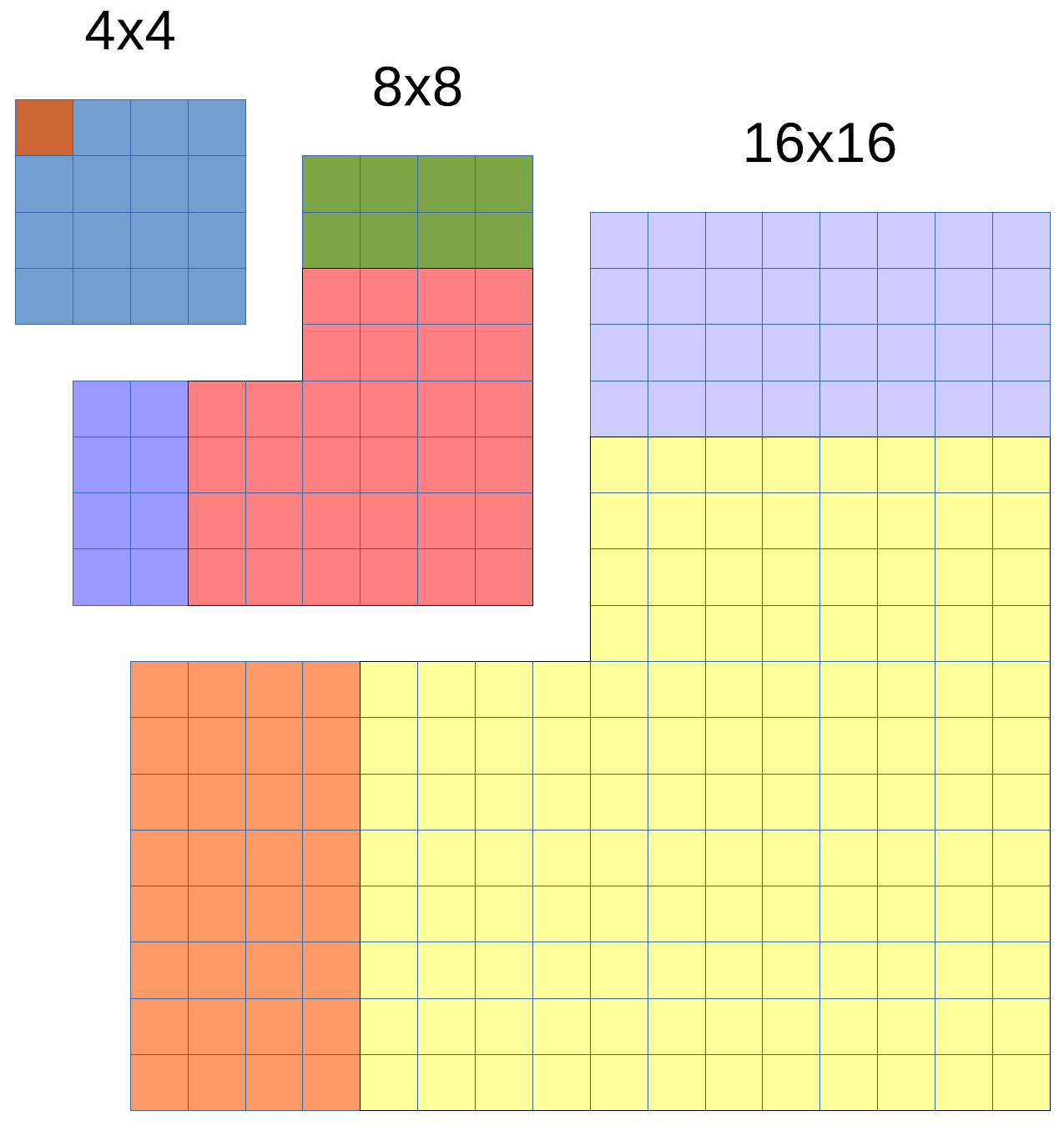}
\end{tabular}
\end{center}
\caption[example]{\label{fig:bands} The band structure of 4x4, 8x8 and 16x16
 blocks in Daala.}
\end{figure}

Up to this point we have only examined the case when all of the AC
 coefficients for an $N\times N$ block are considered together as a single
 input vector for PVQ prediction.
In practice, it may be better to consider portions of the AC coefficients
 together so partitions of the block where $\hat{g}=0$ or $\hat{\theta}=0$ are
 coded more efficiently.
Consider the frequency band structure currently used by Daala in
 Figure \ref{fig:bands}.
The PVQ-CfL technique in the previous section is trivially modified to work
 with any arbitrary partitioning of block coefficients into bands.

Instead of considering whether to flip the direction of ${\bf x}_L$ for each band
 partition individually (a signaling cost of 7 bits per 16x16 block), simply
 look at the lowest 4x4 AC partition and use the flip decision there for the
 entire block.
The assumption is that having those larger low frequency coefficients predicted
 well is far more important than getting it exactly right at higher frequencies.
When the quantization step size is large, the high frequency coefficients will
 be sent to zero regardless.

\section{EXPERIMENTAL EVALUATION}

The two frequency-domain intra-prediction techniques described in this paper for
 predicting chroma coefficients from reconstructed luma coefficients were
 evaluated within the framework of the experimental Daala video codec.
To make the comparison fair, only the AC chroma coefficients were predicted
 using chroma-from-luma with the DC chroma coefficients being predicted as the
 average of its neighbors.
Only gain-shape quantization was used to code the transform coefficients,
 with implicitly predicted FD-CfL coefficients being passed as the predictor
 to PVQ in one instance, and the PVQ-CfL algorithm being used in the other.
All other video coding techniques used were identical.

A sample of 50 still images taken from Wikipedia and downsampled to 1 megapixel
 were compressed at varying quantization levels, and the resulting
 rate-distortion curves were computed on both the Cb and Cr chroma planes using
 four different metrics.
The Bjontegaard distance\cite{bdrate2001} was computed to measure the average
 improvement in both Rate (at equivalent quality) and SNR (at equivalent size)
 between the two techniques.
The result of this experiment is shown in Table \ref{tbl:exp}.
 
\begin{table}[h]
\begin{center}
\begin{tabular}{|c|c|c|c|c|}
\hline
\multirow{2}{*}{Metric} & \multicolumn{2}{|c|}{Cb (plane 1)} & \multicolumn{2}{|c|}{Cr (plane 2)} \\
\cline{2-5}
 & $\Delta$ Rate (\%) & $\Delta$ SNR (dB) & $\Delta$ Rate (\%) & $\Delta$ SNR (dB) \\
\hline
PSNR & -1.27280 & 0.05171 & -0.57558 & 0.02941 \\
\hline
PSNR-HVS & -2.67808 & 0.13703 & -1.24695 & 0.07459 \\
\hline
SSIM & -2.51125 & 0.07116 & -1.66549 & 0.05779 \\
\hline
FastSSIM & -3.18398 & 0.06969 & -2.79776 & 0.06737 \\
\hline
\end{tabular}
\end{center}
\caption[example]{\label{tbl:exp} Computation of the Bjontegaard distance
 (improvement) between the two rate-distortion curves, moving from FD-CfL to
 PVQ-CfL, based on four quality metrics.}
\end{table}

\begin{figure}
\begin{center}
\begin{tabular}{c c}
\includegraphics[width=3.25in]{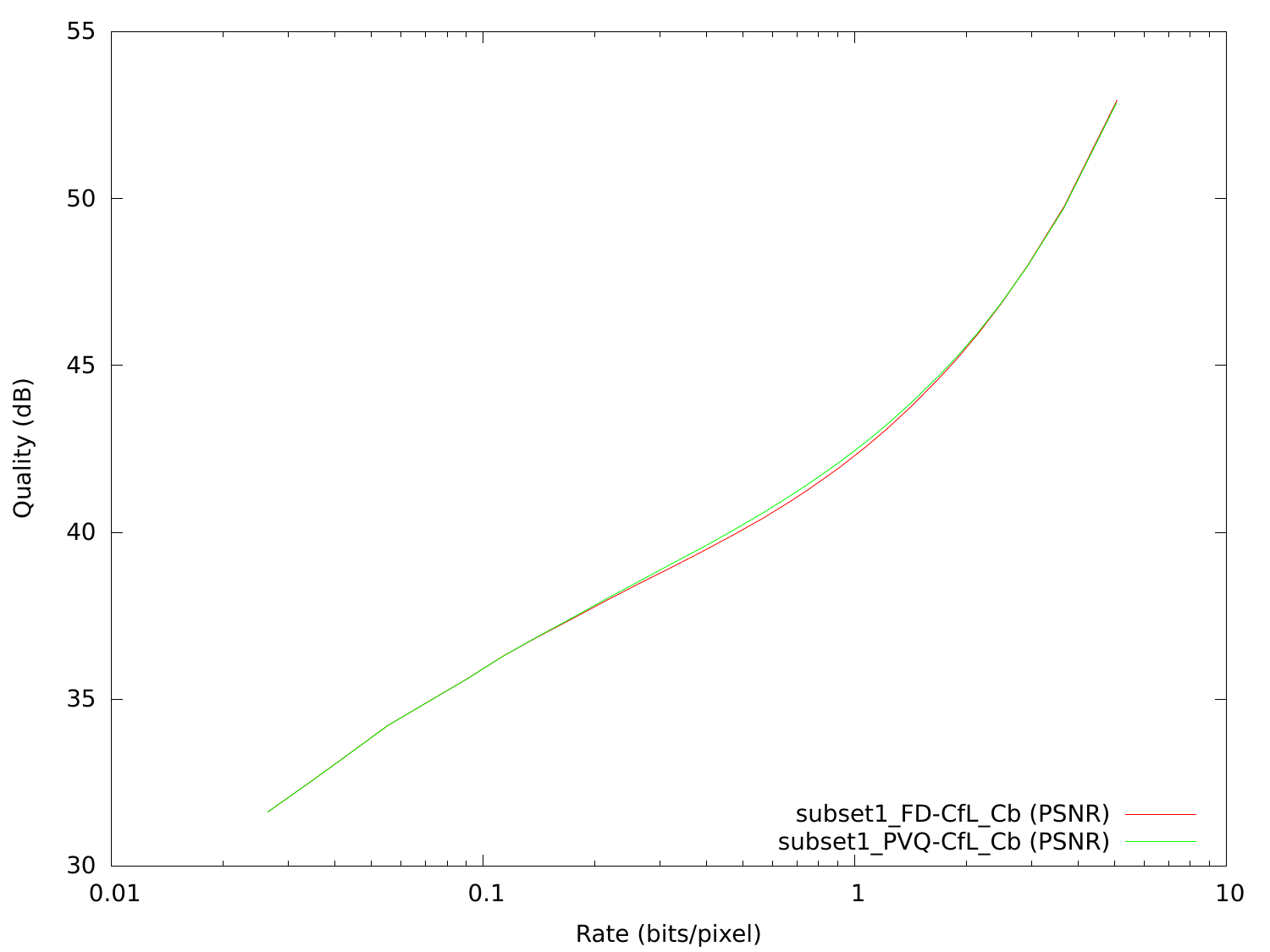} &
\includegraphics[width=3.25in]{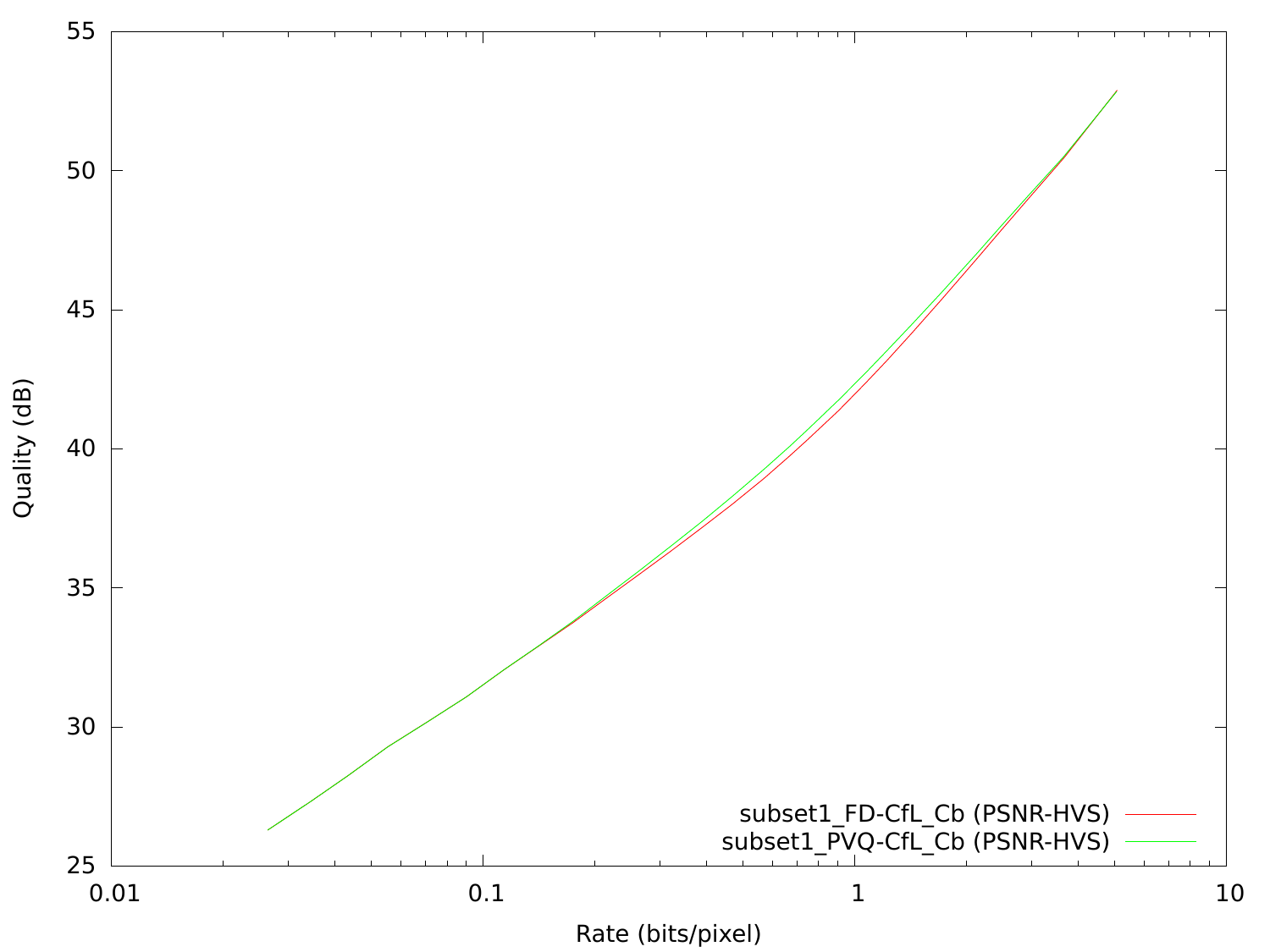} \\
(a) & (b) \\
& \\
\includegraphics[width=3.25in]{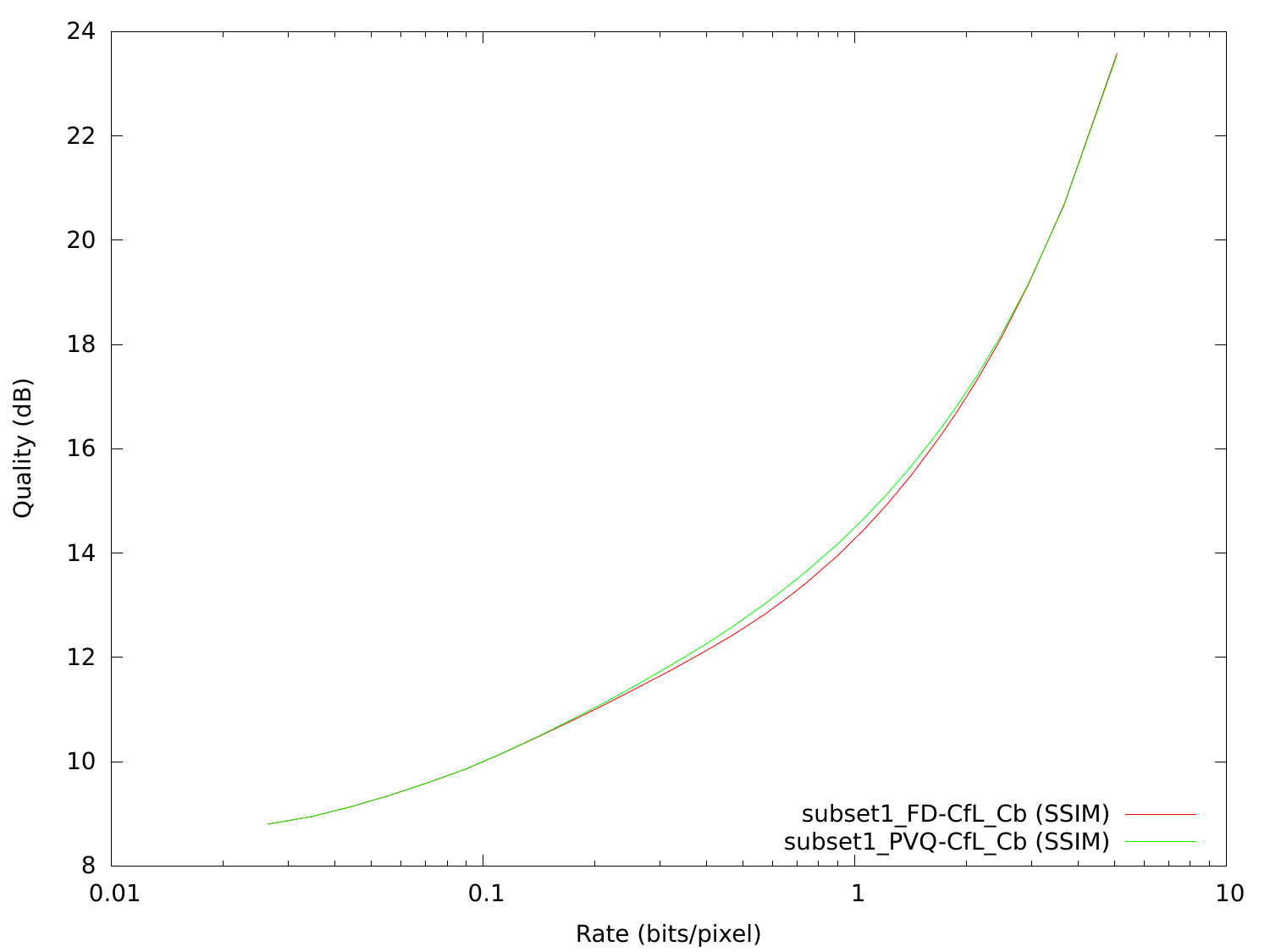} &
\includegraphics[width=3.25in]{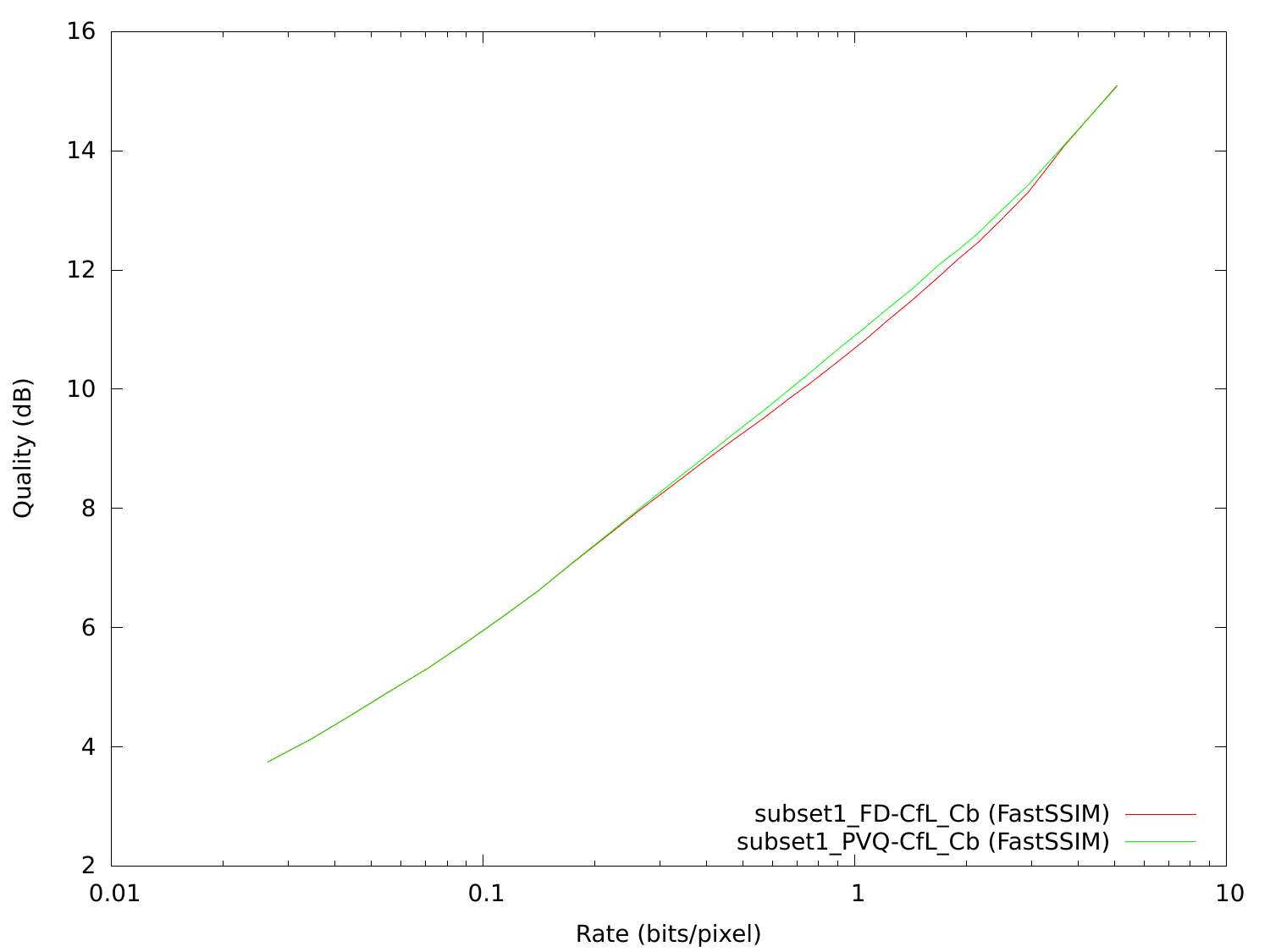} \\
(c) & (d) \\
\end{tabular}
\end{center}
\caption[pvq]{\label{fig:rd} A comparison of the rate-distortion curves for
 FD-CfL and PVQ-CfL on the same set of 50 still images using quality metrics
 (a) PSNR, (b) PSNR-HVS, (c) SSIM and (d) FastSSIM.}
\end{figure}

For every quality metric we considered, the PVQ-CfL technique did a better job
 of predicting Cb and Cr chroma coefficients both delivering higher quality at
 the same rate (up to 0.13 dB improvement for PSNR-HVS\cite{pono2007}) and better compression
 for equivalent distortion (3.2\% smaller files for FastSSIM\cite{Chen2010}).
Looking at the actual rate-distortion curves in Figure \ref{fig:rd}, we see that
 the largest improvements are found at higher rates.
However, the reduced complexity of PVQ-CfL over FD-CfL with no rate or quality
 penalty at lower rates means this technique is clearly superior and thus has been
 adopted in Daala.

\section{CONCLUSIONS}

We have presented two new techniques for doing chroma-from-luma intra-prediction
 in the frequency domain for use with video and still image codecs that employ
 lapped transforms.
The first technique (FD-CfL) is suitable for use in codecs based on scalar
 quantization and provides a reduction in complexity when compared to spatial
 domain CfL as per block complexity stays constant with block size increases.
The second technique (PVQ-CfL) extends the PVQ prediction technique in codecs
 using gain-shape quantization to allow for better chroma prediction with no
 additional per block complexity from model fitting.

This work is part of the Daala project\cite{DaalaWebsite}.
The full source code, including both of the algorithms described in this paper
 is available in the project git repository\cite{DaalaGit}.



\bibliography{spie_cfl}   
\bibliographystyle{spiebib}   

\end{document}